\documentclass[letterpaper,english,reprint,superscriptaddress,longbibliography]{revtex4-1}
\usepackage{lmodern}

\usepackage[T1]{fontenc}
\usepackage[utf8]{inputenc}
\usepackage{babel}
\usepackage{booktabs}
\usepackage{amsmath}
\usepackage{amssymb}
\usepackage{graphicx}
\usepackage[unicode=true,pdfusetitle,
 bookmarks=false,
 breaklinks=false,pdfborder={0 0 0},pdfborderstyle={},backref=false,colorlinks=false]
 {hyperref}

\makeatletter

\pdfpageheight\paperheight
\pdfpagewidth\paperwidth

\providecommand{\tabularnewline}{\\}

\makeatother

\begin{document}
\title{Cross-polarization extinction enhancement and spin-orbit coupling
of light for quantum-dot cavity-QED spectroscopy}
\author{P. Steindl}
\email{steindl@physics.leidenuniv.nl}

\affiliation{Huygens-Kamerlingh Onnes Laboratory, Leiden University, P.O. Box 9504,
2300 RA Leiden, The Netherlands}
\author{J.A. Frey}
\affiliation{Department of Physics, University of California, Santa Barbara, California
93106, USA}
\author{J. Norman}
\affiliation{Department of Electrical \& Computer Engineering, University of California,
Santa Barbara, California 93106, USA}
\author{J.E. Bowers}
\affiliation{Department of Electrical \& Computer Engineering, University of California,
Santa Barbara, California 93106, USA}
\author{D. Bouwmeester}
\affiliation{Huygens-Kamerlingh Onnes Laboratory, Leiden University, P.O. Box 9504,
2300 RA Leiden, The Netherlands}
\affiliation{Department of Physics, University of California, Santa Barbara, California
93106, USA}
\author{W. Löffler}
\email{loeffler@physics.leidenuniv.nl}

\affiliation{Huygens-Kamerlingh Onnes Laboratory, Leiden University, P.O. Box 9504,
2300 RA Leiden, The Netherlands}
\begin{abstract}
Resonant laser spectroscopy is essential for the characterization,
operation, and manipulation of single quantum systems such as semiconductor
quantum dots. The separation of the weak resonance fluorescence from
the excitation laser is key for high-quality single- and entangled
photon sources. This is often achieved by cross-polarization laser
extinction, which is limited by the quality of the optical elements.
Recently, it was discovered that Fresnel-reflection birefringence
in combination with single-mode filtering counteracting spin-orbit
coupling effects enables a three-order of magnitude improvement of
polarization extinction {[}PRX 11, 021007 (2021){]}. Here, we first
investigate multiple reflections and analyze beam reshaping, and observe
that the single-reflection extinction enhancement is optimal. We then
demonstrate this method for cross-polarization extinction enhancement
for a resonantly excited semiconductor quantum dot in a birefringent
optical micro cavity, and observe a 10$\times$ improvement of single-photon
contrast.
\end{abstract}
\maketitle
Highly pure and indistinguishable single-photon sources \cite{Santori2002,Strauf2007,Ding2016,He2013,Somaschi2016,Snijders2018,Tomm2021}
and spin manipulation \cite{Latta2009} often rely on resonant excitation
schemes and require a high degree of polarization extinction  of
the excitation laser. Such experiments are typically carried out in
confocal microscope setups having the required high spatial resolution
\cite{Vamivakas2009,Kaldewey2018} to address individual quantum emitters,
where the excitation beam is directed onto the sample using mirrors
and beamsplitters. Because all-dielectric (but also metallic) polarization-preserving
mirrors \cite{Rudakova2018} do not exist, usually, linearly polarized
light with $s$ or $p$ polarization is used, which, due to symmetry
and in the plane-wave approximation, is preserved under reflection.
Therefore, the maximal polarization extinction ratios (PER) are limited
by the quality of available polarizers to typically $10^{5}-10^{6}$.
Now, experimentally, extinction ratios of up to $10^{8}$ \cite{Kuhlmann2013}
have been observed, but the precise origin of this high ratio has
only recently been clearly identified. Benelajla et al. \cite{Benelajla2021}
found a 3-orders of magnitude PER improvement by mirror-induced pre-compensation
of the residual ellipticity of linear polarizers, in combination with
single-mode filtering that eliminates detrimental effects caused by
spin-orbit coupling at optical reflection \cite{Gotte2012,Bliokh2013,Toppel2013}. 

Spin-orbit coupling of light leads to angular and spatial beam shift
corrections to specular reflection \cite{Merano2009,Aiello2009},
known as Goos–Hänchen \cite{Goos1947} and Imbert–Fedorov shifts \cite{Imbert1972,Fedorov2013}.
The latter effect is also referred to as the optical spin-Hall effect
of light \cite{KWIAT2008,Bliokh2015}. Here, both of the elliptical
eigenpolarizations \cite{Gotte2014} experience a small opposite transverse
shift which leads to a variation in the degree of the circular polarization
over the beam cross-section \cite{Bliokh2006}. Measuring in linear
cross-polarization, the unwanted (leaked) polarization component of
the beam adopts a Hermit-Gaussian profile \cite{Xiao2011,Jayaswal2014,Benelajla2021}
with a nodal line in the center, caused by linear polarization projection
of the two reflected and shifted beams with opposite helicity. In
combination with single-mode fiber detection, Benelajla et al. \cite{Benelajla2021}
demonstrated above three orders extinction enhancement after a single
optical reflection compared to the bare or conventional polarizer
extinction ratio with $90^{\circ}$ between the polarizers. This shows
a new - only used serendipitously before - application of beam-shifts
in addition to high-resolution sensing \cite{Yin2006,Ling2017} and
corrections in astrophysical instruments \cite{Schmid2018}, but the
effect has not been clearly identified in confocal microscopy.

Here, we first demonstrate that the elimination of residual scattering
from a Glan-type polarizer by propagation or single-mode fiber filtering
already enables extinction ratios beyond $10^{7}$, then we investigate
the effect of multiple in-plane reflections (under $45^{\circ}$ angle
of incidence) on the PER. We find that a single Fresnel reflection
is optimal, achieving a PER of nearly $10^{8}$. Finally, we demonstrate
the PER enhancement effect by reflection in a cryogenic confocal microscope.
By careful characterization and optimization of the polarizer rotation
angles, we achieve extinction ratios up to $10^{7}$, two orders above
the bare polarizer limit, and can clearly attribute this effect to
ellipticity and spin-orbit coupling compensation. With this, we demonstrate
an improvement of a quantum-dot cavity-QED-based single-photon source.

\section{scattering elimination\label{sec:scattering-elimination}}

\begin{figure}[h]
\includegraphics[width=1\columnwidth]{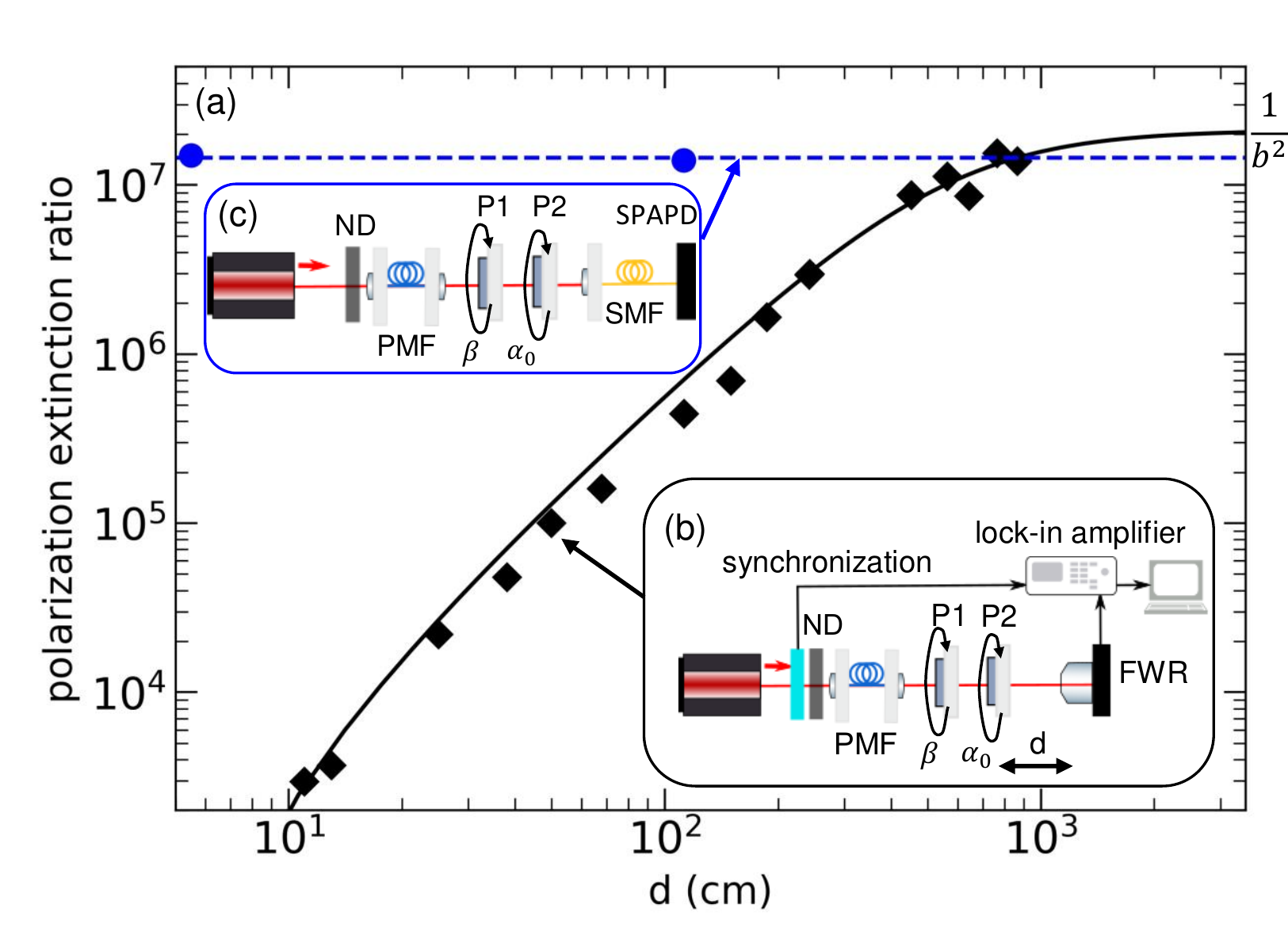}

\caption{The effect of scattered light on the cross-polarization extinction
ratio. (a) black symbols: free-space detected PER for increasing detector
distance $d$ from the analyzing polarizer, the experimental configuration
is sketched in (b). blue: extinction measured using single-mode fiber
filtering, (c) shows the setup. The black curve is a model fit as
described in the main text. \label{FIG:Scattering_elimination}}
\end{figure}

We start with a simple optical setup shown in the insets of Fig. \ref{FIG:Scattering_elimination}.
A narrow-linewidth continuous-wave laser with wavelength $\lambda=935.5\,\text{nm}$
is attenuated with a calibrated set of neutral density filters, mode-filtered
with a polarization-maintaining single-mode fiber (PMF) and collimated
by an aspheric lens into a Gaussian beam with 0.75 mm beam waist.
It is sent to the two polarizers P1 (which is approximately aligned
to the incident-light polarization) and P2, which are here two anti-reflection
coated calcite Glan-Thompson polarizers, placed 15 cm from each other
and mounted in motorized rotation stages with $10^{-2}$ and $10^{-3}$
degree resolution, respectively. To analyze the polarizer extinction,
the light transmitted through both polarizers is focussed with a lens
onto a femtowatt photo receiver (FWR) that is placed at an adjustable
distance $d$ from the analyzer P2, see Fig. \ref{FIG:Scattering_elimination}(c).
To subtract background light, we use laser modulation and a lock-in
amplifier. We determine the polarization extinction ratio $\text{PER}=I_{\text{copol}}/I_{\text{xpol}}$
from the co- and cross-polarized transmitted intensities $I_{\text{copol}}$
and $I_{\text{xpol}}$. The front polarizer is kept at a fixed angle
$\beta$ aligned roughly to the polarization of the laser source and
$I_{\text{xpol}}$ is minimized by rotation of P2.

We measure PER for various distances $d$ up to 8.5 m, collecting
in all cases over 90\% of the beam area. The data presented in Fig.
\ref{FIG:Scattering_elimination}(a) shows a gradual increase of the
measured PER with distance, ranging from $10^{3}$ at few centimeter
distances to $1.4\times10^{7}$ at 8.5 m. Our observed dependency
on distance can be explained by scattering including Rayleigh scattering
\cite{Moeller1969}: We model the unwanted light in cross-polarization
by (i) residual ellipticity of the polarizers \cite{King1971} that
is limiting the bare extinction ratio to $1/b^{2}$ \cite{Benelajla2021},
and (ii) (Rayleigh) scattered light, which is quadratically decreasing
with distance \cite{Cox2002}, resulting in $I_{\text{xpol}}=b^{2}+b_{\text{scat}}^{2}/d^{2}$.
We fit our data and obtain\textbf{ $b=(2.1\pm0.2)\times10^{-4}$ }and\textbf{
$b_{\text{scat}}=(12.6\pm1.2)\times10^{-5}\,\mathrm{m}^{-1}$}, limiting
the bare extinction measured after scattering elimination to $2.1\times10^{7}$.
This value is very close to the extinction of $1.4\times10^{7}$ that
we achieve by single-mode fiber (SMF) filtering shown in Fig. \ref{FIG:Scattering_elimination}(c).
We repeat the SMF-filtered experiment at two distances from P2, including
at $d=4.5$ cm where scattering is dominant without fiber filtering,
and observe a constant PER. As expected, a single-mode fiber efficiently
removes scattered light.

We operate our photodetector close to the dark current limit, and
the signal fluctuated by 5\% for the highest PERs. Therefore we repeated
the experiments with fiber filtering with a fiber-coupled single-photon
avalanche photon detector (SPAPD, 25 \% detection efficiency, $\ensuremath{200\,\mathrm{s}^{-1}}$dark
count rate), where the measured PER is limited by dark counts to $6\times10^{11}$.
We obtain a polarization extinction of $1.5\times10^{7}$, confirming
our previous results. This is more than two orders of magnitude higher
than specified. We have also repeated the same experiment with different
pairs of the Glan-Thompson polarizers and always found extinction
ratios above $10^{7}$. This agrees to earlier studies \cite{Moeller1969,King1971},
only surpassed by dedicated studies using $10^{-4}$ degree resolution
rotation stages, resulting in extinction ratios up to $\sim3\times10^{9}$
\cite{Takubo1998,Mei2006,He2016}. In the Appendix, we present additional
measurements for various analyzers, and show the effect of the anti-reflection
coating of Glan-Thompson polarizers, and compare Glan-Thompson polarizers
with nanoparticle polarizers.

\section{Vector-beam effects upon multiple reflections\label{sec:Vector-beam-effects-upon}}

\begin{figure}[h]
\includegraphics[width=1\columnwidth]{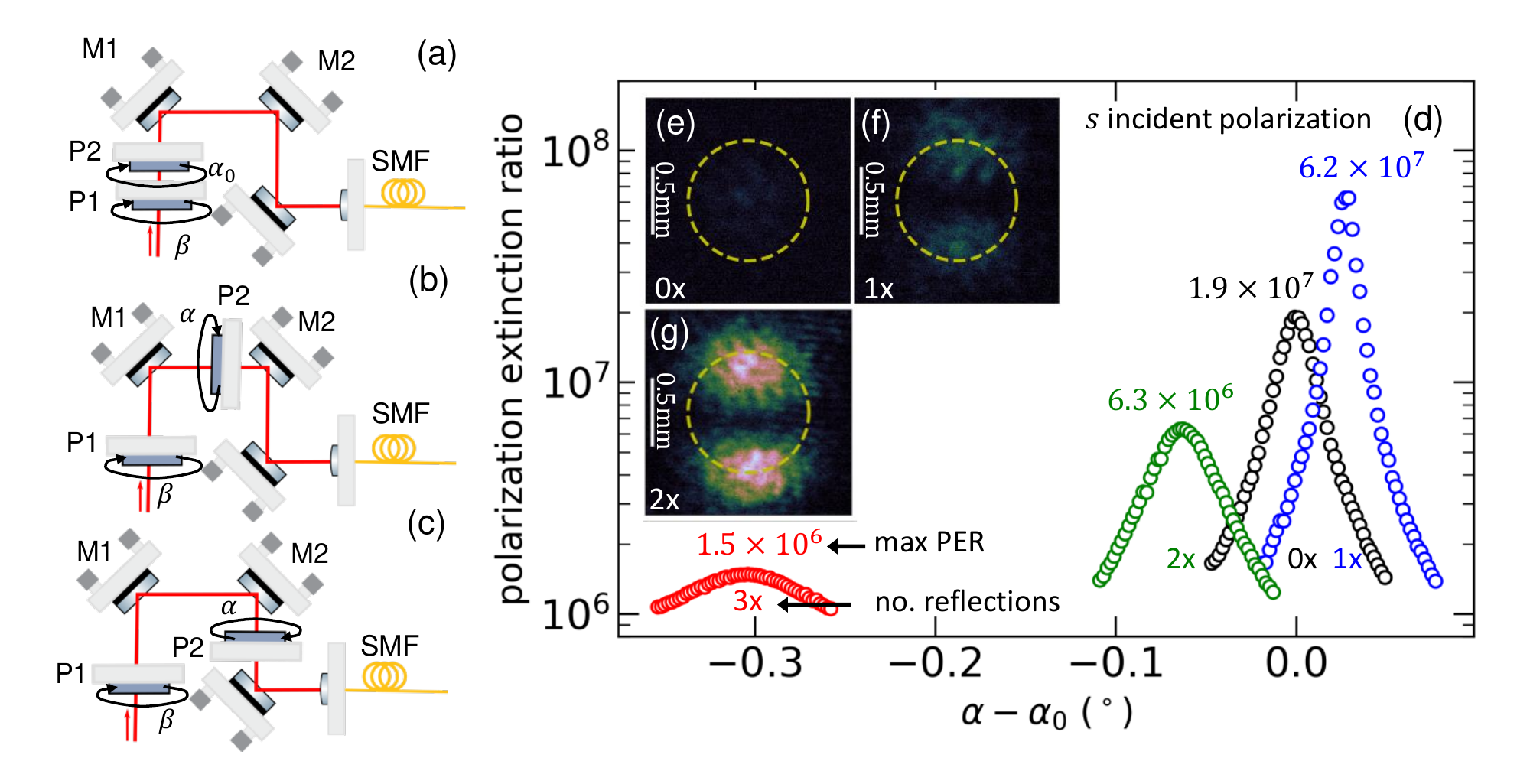}

\caption{Polarization extinction and spin-orbit coupling upon multiple reflections.
We use a series of setups with a different number of optical reflections
between the polarizers (a-c), determine the maximum achievable polarization
extinction (d), and also show the residual unwanted light pattern
measured in cross-polarization before coupling into the single-mode
fiber (e-g, the dashed circle shows the fiber mode before the collimation
lens). In (d), the PER as a function of the analyzing polarizer angle
$\alpha$ relative to the conventional cross-polarization angle $\alpha_{0}=\beta\pm\pi/2$
is shown. The maximal PER for each configuration with $n$ reflections
is indicated.\label{FIG:Multiple-reflections_scans}}
\end{figure}
Now we investigate the influence of (multiple) in-plane reflections
on the achievable polarization extinction ratio using experimental
setups with 0, 1, and 2 reflections between the polarizers, as shown
schematically in panels (a)–(c) of Fig. \ref{FIG:Multiple-reflections_scans}.
All mirrors are dielectric thin-film mirrors placed approximately
under $45^{\circ}$ of incidence, and the last mirror in combination
with a translation of the fiber collimation lens is used to optimize
coupling into the single-mode fiber, where we achieve a coupling efficiency
of 80 \% in all cases. The single-mode fiber removes scattered light
as investigated in the previous section. The initial beam is approximately
$s$-polarized by P1, both P1 and P2 are Glan-Thompson polarizers.
We show similar results for $p$ polarization in the Appendix. For
each configuration, the polarizer angles $\alpha$ and $\beta$ are
fine-adjusted iteratively around the conventional cross-polarization
condition ($\left|\alpha_{0}-\beta\right|=\pi/2$) to optimize the
PER. Note that even a small deviation of about $0.03^{\circ}$ from
the optimal settings of P1 leads to a reduction of PER by one order
of magnitude.

First, we discuss the single-reflection case as shown in Fig. \ref{FIG:Multiple-reflections_scans}(b),
where we reliably reach an extinction ratio of $6.2\times10^{7}$.
This extinction ratio is by a factor 3 higher compared to the zero-reflection
configuration shown in Fig. \ref{FIG:Multiple-reflections_scans}(a).
The enhancement is due to compensation of the residual polarizer ellipticities
by the small Fresnel-reflection birefringence if the light incident
on the mirror is not exactly $s$ or $p$ polarized; and possibly
a stress-induced birefringence of the mirror coating. In any case,
this birefringence enables for compensation of residual ellipticities
of the used polarizers and thus improves the PER \cite{Benelajla2021}
compared to the zero-reflection case. As mentioned by Benelajla et
al. \cite{Benelajla2021}, this can be explained in the plane-wave
picture because vector-beam effects and spin-orbit coupling results
in higher-order modes, in our case mainly in the first-order Hermite-Gaussian
mode \cite{Aiello2007,A.Aiello2008,Benelajla2021,Jayaswal2014}. This
mode has a nodal line in the center and is shown in Fig. \ref{FIG:Multiple-reflections_scans}(f),
and it can largely be filtered out by the detection single-mode fiber
and does therefore not degrade much the PER.

To further investigate this argument, we have also tested two {[}Fig.
\ref{FIG:Multiple-reflections_scans}(c){]} and three reflections,
each time optimized the polarizer angles, and we observe a reduced
PER compared to a single reflection, $6\times10^{6}$ after two reflections
and $1\times10^{6}$ after three reflections. We also observe an increase
of the intensity in the first-order Hermite-Gauss mode as shown in
Fig. \ref{FIG:Multiple-reflections_scans}(g), suggesting that imperfect
single-mode filtering can explain the reduction of the PER for multiple
reflections.

\section{Single emitter polarization extinction improvement\label{sec:Polarization-extinction-improvem}}

\begin{figure}[h]
\includegraphics[width=1\columnwidth]{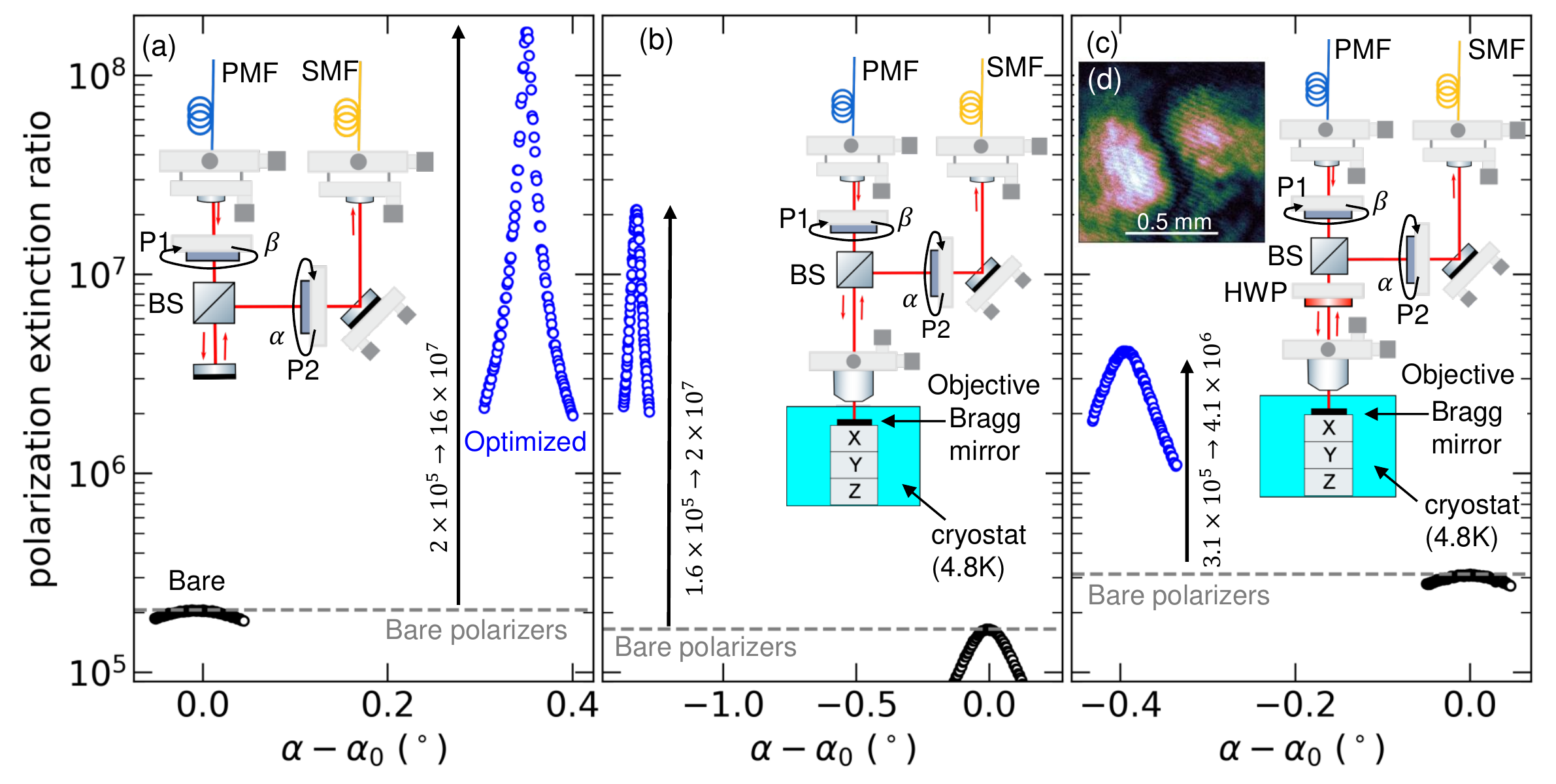}

\caption{Polarization extinction improvement (blue symbols) by reflection in
a nonpolarizing beamsplitter in a cryogenic microscope. Shown is the
PER as a function of the angle of the analyzing polarizer $\alpha$
relative to the conventional cross-polarization angle $\alpha_{0}$.
First, the case is shown where the cryostat is replaced by a flat
mirror, where beyond $10^{8}$ improvement is obtained (a). Then the
beam is focussed into the cryostat and reflected there and we obtain
beyond $10^{7}$ contrast (b). Finally, the case is shown where a
half-wave plate is added before the objective in order to align the
linearly polarized cavity modes of the device in the cryostat to the
polarization frame of the optical setup (c). \label{FIG:PER-confocal_microscope}}
\end{figure}
Now we investigate whether the method to improve the polarization
extinction ratio can also be applied in more complex experimental
setups, for which we investigate resonant optical spectroscopy of
a single self-assembled InGaAs/GaAs quantum dot (QD) embedded at the
antinode of a high-quality micropillar cavity \cite{Snijders2018,Steindl2021_PRL},
using various setups shown in the insets of Fig. \ref{FIG:PER-confocal_microscope}.
Instead of the mirror reflection and to independently vary the excitation
(P1) and detection polarization (P2), we use a non-polarizing beamsplitter
(BS) with 90:10 (R:T) splitting ratio to separate excitation and detection
paths. The polarization of the excitation narrow-linewidth laser light
($\lambda=935.5\,\text{nm}$, $\text{FWHM}=200\,\mathrm{kHz}$, beam
waist $0.75\,\mathrm{mm}$) is controlled with a Glan-Thompson polarizer
mounted in a rotation stage with an angular resolution of $0.01\,\mathrm{deg}$.
The detection polarization selection is done with a nanoparticle polarizer
which is less sensitive to alignment, mounted in a $0.001\,\mathrm{deg}$
resolution motorized stage; this polarizer has a bare extinction of
$1.9\times10^{5}$. The transmitted light is coupled into a single-mode
fiber (coupling efficiency \textasciitilde 85\%) and detected with
a SPAPD. 

First, we reflect the incident light from a plane dielectric mirror
under normal incidence placed below the beam splitter, as sketched
in Fig. \ref{FIG:PER-confocal_microscope}(a). We optimize the PER
close to $s$ polarization (at the beam splitter) and obtain a PER
of $1.6\times10^{8}$. This is nearly three orders of magnitude higher
compared to bare extinction measured with the nanoparticle polarizer
in conventional cross-polarization, reproducing our previous results.

Now, we use a long-working-distance plan apochromat objective (0.4
NA, infinity corrected) focussing the light through two silica glass
windows into a close-cycle cryostat cooled to $4.8\,\mathrm{K}$.
The light is reflected from the top GaAs/AlAs thin-film Bragg mirror
of our quantum dot-cavity device. We repeat the optimization of the
polarizer angles which are different due to polarization changes caused
by the objective and the two silica windows. As shown in Fig. \ref{FIG:PER-confocal_microscope}(b),
even in this complex configuration, we reach an extinction ratio of
$2\times10^{7}$, a factor 100 above the bare polarization extinction
ratio. 

Now we perform single-emitter spectroscopy of a single quantum dot
in a Fabry-Perot microcavity at around 935.5 nm. The fundamental cavity
mode is split by shape and strain induced birefringence \cite{Frey2018,Snijders2020}
by $\sim28\,\mathrm{GHz}$ into two orthogonal linearly polarized
modes ($V$ and $H$). In order to align the polarization frame of
reference of the confocal microscope to the frame of the cavity, while
avoiding the need to rotate either of them, we use an extra half-waveplate
(HWP) below the beamsplitter as shown in Fig. \ref{FIG:PER-confocal_microscope}(c).
We again optimize the angles of P1 and P2 and reach a PER of $4\times10^{6}$
away from the cavity resonances, this is still more than one order
of magnitude higher than the bare polarizer extinction ratio. At the
same time, as shown in Fig. \ref{FIG:PER-confocal_microscope}(d),
we observe in cross-polarization again the typical Hermite-Gauss mode.
This suggests that we have demonstrated polarization extinction improvement
in a complex cryogenic confocal microscope, despite detrimental effects
of the microscope objective and of the focused beam through two cryostat
windows, reaching similar extinction ratios as with in-cryostat focussing
optics \cite{Kaldewey2018}.

\begin{figure}[h]
\includegraphics[width=1\columnwidth]{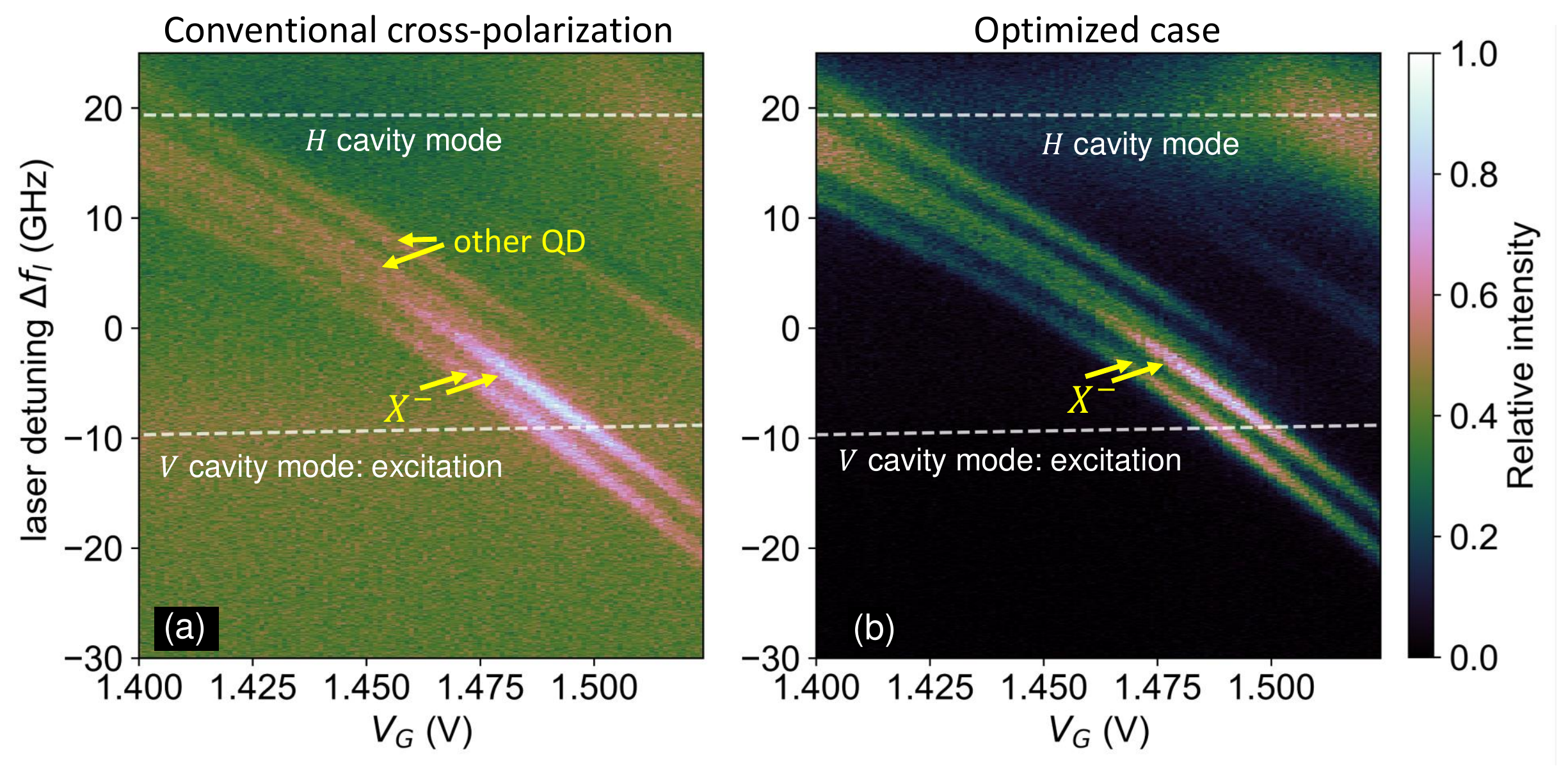}

\caption{Resonant cross-polarized single quantum dot spectroscopy for the conventional
cross-polarization configuration (a) and the optimized case (b). Shown
is the cross-polarized resonance fluorescence of a negatively charged
trion tuned through the cavity resonances by the quantum-confined
Stark effect, as a function of laser detuning $\Delta f_{l}$ and
gate voltage $V_{G}$. The incident laser light is polarized along
the $V$ cavity axis. Dashed lines indicate both cavity resonance
frequencies determined by fits of a semiclassical model \cite{Snijders2020}.
\label{FIG:Voltscan_trion}}
\end{figure}

Finally, we demonstrate that the method is also compatible with GHz-scale
tuning of the laser, we now show resonant spectroscopy of the negatively
charged trion transition $X^{-}$ of the quantum dot. The dot is embedded
in the intrinsic region of a \textit{p-i-n} diode that allows Stark-shift
tuning of the quantum dot resonance through the linearly polarized
cavity resonances, see Refs. \cite{Frey2018,Snijders2020}. Under
an in-plane (Voigt geometry) magnetic field of $500\,\mathrm{mT}$
we observe in Fig. \ref{FIG:Voltscan_trion} the expected transition
doublet \cite{Bayer2002}; each plot shows the normalized single-photon
detection rate. Compared to the conventional cross-polarization condition
shown in Fig. \ref{FIG:Voltscan_trion}(a) we see a clear improvement
by the optimization polarization condition in Fig. \ref{FIG:Voltscan_trion}(b).
The ratio of the single photons emitted by the quantum dot to background
laser light increases from \textasciitilde 2 to \textasciitilde 25
- a significant improvement if used as a single-photon source. 

\section{Conclusions}

In conclusion, we have clearly identified the polarization extinction
improvement effect by optical reflection and spin-orbit coupling of
Benelajla et al. \cite{Benelajla2021} in a complex cryogenic confocal
microscope setup and obtained more than one order of magnitude improvement
of a quantum-dot based single-photon source. We have shown that single-mode
fiber detection is important for the reduction of both unwanted scattered
light as well as higher-order modes that appear unavoidably by spin-orbit
coupling at the mirror or beamsplitter. Only with this mode filtering,
the polarization extinction ratio enhancement can be described in
a plane-wave picture as a reflection-induced birefringence compensation
of the residual elliptical polarization of linear polarizers. We have
demonstrated that this extinction enhancement has a direct impact
on the single-photon contrast in resonant quantum dot spectroscopy,
which we have demonstrated with a single quantum dot in a polarization
non-degenerate optical micro cavity. To obtain even higher polarization
extinction ratios than those reported here ($4\times10^{6}$), one
should rotate the cavity device and remove the half-wave plate since
this wave plate most likely also leads to spin-orbit coupling induced
modal changes \cite{BLIOKH2016} similar to multiple reflections that
we have investigated here.
\begin{acknowledgments}
We thank Harmen van der Meer and Martin van Exter for discussions
and support. We acknowledge for funding from the European Union’s
Horizon 2020 research and innovation programme under grant agreement
No. 862035 (QLUSTER), from FOM-NWO (08QIP6-2), from NWO/OCW as part
of the Frontiers of Nanoscience program and the Quantum Software Consortium,
and from the National Science Foundation (NSF) (0901886, 0960331).
\end{acknowledgments}


\pagebreak
\renewcommand{\thefigure}{A\arabic{figure}}\setcounter{figure}{0}\renewcommand{\theequation}{A\arabic{equation}}\setcounter{equation}{0}\renewcommand{\thetable}{A\arabic{table}}\setcounter{table}{0}\appendix

\section{Scattering elimination with various analyzers}

\begin{figure}[h]
\includegraphics[width=1\columnwidth]{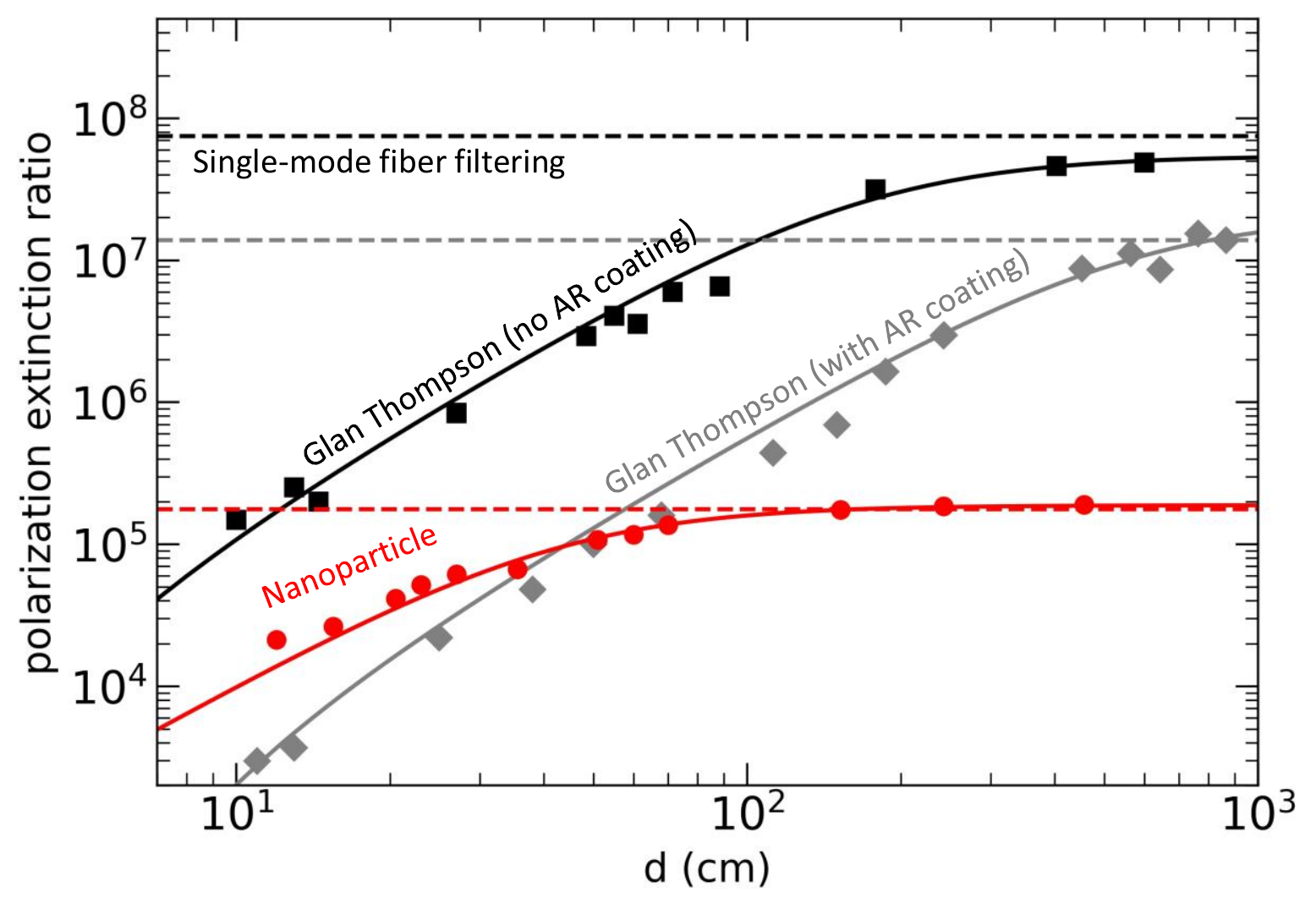}

\caption{Maximum conventional (without mirror reflection) polarization extinction
ratio for various analyzer types and conditions: The dashed horizontal
lines show the PER using a single-mode fiber, the symbols show the
PER using a free-space photo detector at various distances $d$ behind
P2. As analyzing polarizer, we used a nanoparticle polarizer (red),
and a Glan-Thompson polarizer without (black) and with (gray) anti-reflection
coating. Solid lines show the model fit. \label{FIG:Scattering_elimination-various_analyzers}}
\end{figure}

In Sec. \ref{sec:scattering-elimination} of the main text, we have
discussed the importance of the elimination of scattered light for
maximal PER. Here we show results (Fig. \ref{FIG:Scattering_elimination-various_analyzers})
for a wider range of analyzing polarizers P2. We compare Glan-Thompson
analyzing polarizers with and without anti-reflection (AR) coating,
and a nanoparticle polarizer; the polarizer P1 is an AR-coated Glan-type
polarizer in all cases. 
\begin{table*}
\begin{centering}
{\footnotesize{}}%
\begin{tabular*}{14.5cm}{@{\extracolsep{\fill}}ccccc}
\toprule 
{\scriptsize{}Analyzer type} & {\scriptsize{}$b$ $(10^{-4})$} & {\scriptsize{}$b_{\text{scat}}$$(10^{-4}m^{-1})$} & {\scriptsize{}Bare extinction $1/b^{2}$} & {\scriptsize{}SMF measured PER}\tabularnewline
\midrule
\midrule 
{\scriptsize{}Glan Thompson polarizer (wo AR coating)} & {\scriptsize{}$1.35\pm0.03$} & {\scriptsize{}$2.3\pm0.1$} & {\scriptsize{}$(5.45\pm0.2)\times10^{7}$} & {\scriptsize{}$7.50\times10^{7}$}\tabularnewline
\midrule 
{\scriptsize{}Glan Thompson polarizer (with AR coating)} & {\scriptsize{}$2.1\pm0.2$} & {\scriptsize{}$12.6\pm1.2$} & {\scriptsize{}$(2.1\pm0.3)\times10^{7}$} & {\scriptsize{}$1.38\times10^{7}$}\tabularnewline
\midrule 
{\scriptsize{}Nanoparticle polarizer} & {\scriptsize{}$23.0\pm0.3$} & {\scriptsize{}$9.8\pm0.3$} & {\scriptsize{}$(1.88\pm0.04)\times10^{5}$} & {\scriptsize{}$1.76\times10^{5}$}\tabularnewline
\bottomrule
\end{tabular*}{\footnotesize\par}
\par\end{centering}
\caption{Fit parameters of the measurements shown in Fig. \ref{FIG:Scattering_elimination-various_analyzers}.
\label{Tab:Rayleigh_scattering}}
\end{table*}

As in the main text, we model the distance-dependent PER and fit to
the data, and we show the fit results of the cross-polarization intensity
$I_{\text{xpol}}=b^{2}+b_{\text{scat}}^{2}/d^{2}$ in Table \ref{Tab:Rayleigh_scattering}.
As expected \cite{King1971}, our data shows that AR coating leads
to (i) additional scattering captured as an increase in $b_{\text{scat}}$
and (ii) extra residual ellipticity of the polarizer reducing the
maximal bare extinction. The maximal bare extinction ratio achieved
with the nanoparticle polarizer is strongly reduced, most likely due
to residual birefringence of its glass support \cite{King1971}.

\section{Maximal polarization extinction upon multiple reflections}

\renewcommand{\thefigure}{B\arabic{figure}}\setcounter{figure}{0}\renewcommand{\theequation}{B\arabic{equation}}\setcounter{equation}{0}\renewcommand{\thetable}{B\arabic{table}}\setcounter{table}{0}

\begin{figure}[h]
\includegraphics[width=1\columnwidth]{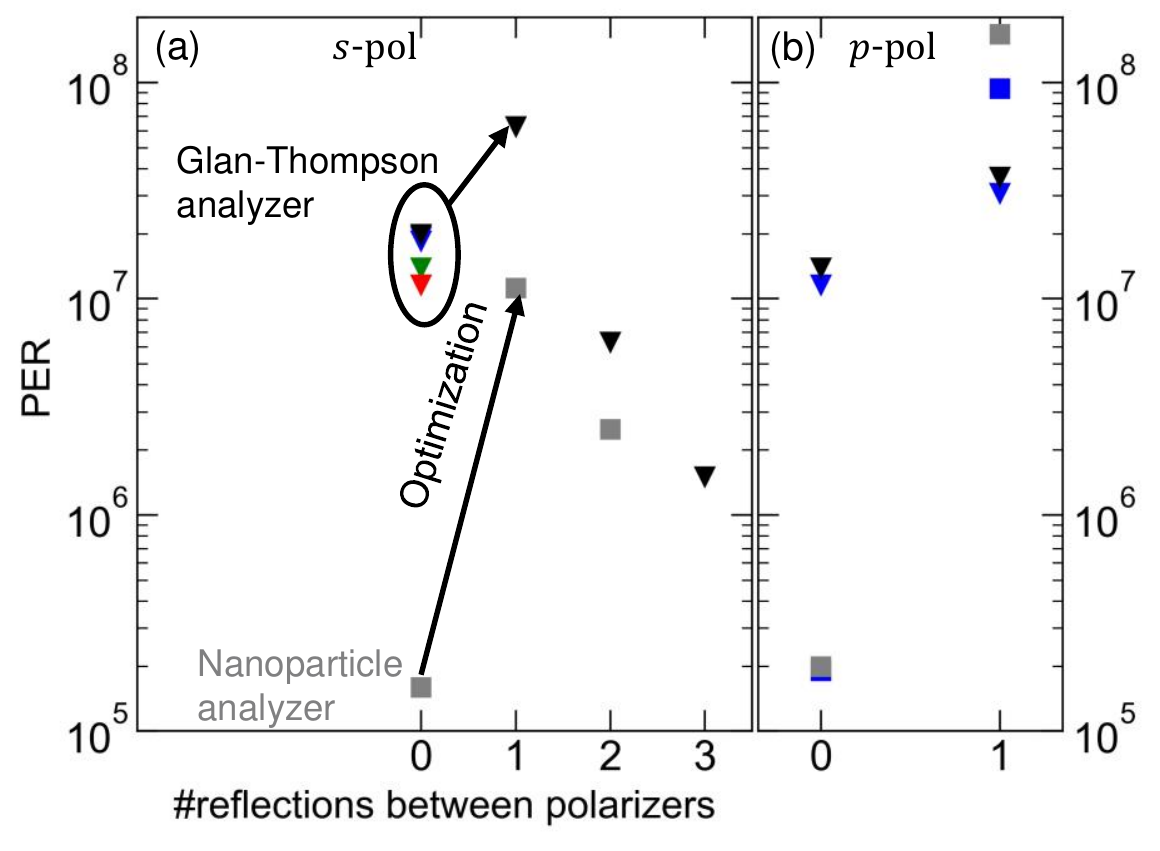}

\caption{Polarization extinction ratio for a different number of reflections
for $s$-polarized (a) and $p$-polarized (b) incident light, for
a Glan-Thompson analyzer (triangles) and a nanoparticle analyzer (squares).
\label{FIG:PER_vs_noreflections_sandp}}
\end{figure}

In this section we show additional data for multiple reflections for
different polarizer types, and also for $s$-polarization. As reported
by Benelajla et al. \cite{Benelajla2021} and shown in Fig. \ref{FIG:PER_vs_noreflections_sandp},
this PER improvement is achieved for both $s$- and $p$-polarized
light. While the observation of the enhancement is not polarizer-type
specific, the absolute value of the maximal PER is a function of the
detrimental ellipticity of the polarizers, thus it can vary between
individual polarizers.

\end{document}